\documentclass[aps,prb,twocolumn,groupedaddress,showpacs]{revtex4-1}

\usepackage{graphicx}
\usepackage{color}
\usepackage{ulem}
\usepackage{verbatim}
\usepackage{hyperref}
\usepackage{amsmath,amssymb}

\begin{document}

\title{The paraconductivity of granular Al-films at high reduced temperatures and magnetic fields}

\author{D. S\'o\~nora$^1$}
\author{C. Carballeira$^1$}
\author{J.J. Ponte$^2$}
\author{F. Vidal$^1$}
\author{T. Grenet$^3$}
\author{J. Mosqueira$^1$}

\email[]{j.mosqueira@usc.es}

\affiliation{$^1$Quantum Materials and Photonics Research Group, Departamento de F\'isica de Part\'iculas, Universidade de Santiago de Compostela, E-15782 Santiago de Compostela, Spain}

\affiliation{$^2$Unidade de Magnetosusceptibilidade, RIAIDT, Universidade de Santiago de Compostela, E-15782 Santiago de Compostela, Spain}

\affiliation{$^3$Institut N\'eel, CNRS and Universit\'e Joseph Fourier, B.P. 166, F-38042 Grenoble C\'edex 9, France}

\date{\today}

\begin{abstract}
The electrical conductivity induced near the superconducting transition by thermal fluctuations was measured in different granular aluminum films. The seemingly anomalous behavior at high reduced temperatures and magnetic fields is explained by taking into account a total-energy cutoff in the superconducting fluctuation spectrum in both the direct (Aslamazov-Larkin) and the indirect (anomalous Maki-Thompson) contributions to the fluctuation effects. The analysis allowed a reliable determination of the coherence length amplitudes, which resulted to be much larger ($20-48$~nm) than the grains size ($5-10$~nm). This suggests that the grains are strongly Josephson-coupled, while the $T_c$ value is still as high as twice the bulk value. These results could contribute to identify the mechanisms enhancing $T_c$ in these materials. 
\end{abstract}


\maketitle

\section{Introduction}

As summarized in Tinkham's textbook on superconductivity \cite{Tinkham} (see also Ref.~\onlinecite{Skocpol75}), the increase just above $T_c$ of the electrical conductivity relative to the normal-state contribution (the so-called paraconductivity, $\Delta\sigma$) is explained in low-$T_c$ superconductors (LTSC) by taking into account two contributions: a \textit{direct} (Aslamazov-Larkin, AL) contribution, due to Cooper pairs created by thermal agitation, and the \textit{indirect} (Maki-Thompson, MT) contribution, associated with the quasiparticles created by the decay of these fluctuating pairs. However, this scenario does not explain the steep $\Delta\sigma$ reduction at high reduced-temperatures, $\varepsilon\equiv\ln(T/T_c)\stackrel{>}{_\sim}0.1$, or magnetic fields, $h\equiv H/H_{c2}(0)\stackrel{<}{_\sim}1$, $H_{c2}(0)$ being the upper critical field linearly extrapolated to $T=0$~K.\cite{Patton71} These difficulties are not completely overcome with the introduction of a \textit{momentum} (or \textit{kinetic-energy}) cutoff to eliminate the contribution of short-wavelength fluctuations.\cite{Tinkham,Skocpol75,Larkinbook,Glatz11}
Alternative scenarios to explain this behavior have been proposed, particularly for the highly studied granular LTSC films, in terms of the reduction of the fluctuations dimensionality, in turn associated with the decoupling of superconducting grains or to a confinement due to electronic inhomogeneities.\cite{Deutscher77,Carbillet16,Pracht16} Nevertheless, some central aspects, as for instance the role that may play the presence of MT contributions, still remain not well settled.

To address the long standing open problem of the $\Delta\sigma$ behavior at high-$\varepsilon$ and -$h$ in LTSC, in this work we first present detailed $\Delta\sigma(T,H)$ measurements around $T_c$ in disordered aluminum films. Then, these data are quantitatively explained, also in the high-$\varepsilon$ and -$h$ regions, by introducing in the AL and the anomalous MT contributions the so-called \textit{total-energy} cutoff, that takes into account the limits imposed by the Heisenberg uncertainty principle to the shrinkage of the superconducting wavefunction.\cite{Vidal02,notaMT} 
The adequacy of this total-energy cutoff to explain the fluctuation-induced (Schmidt) diamagnetism at high-$\varepsilon$ and $h$ in several LTSC has been probed previously.\cite{Mosqueira01b,Mosqueira03,Soto04,Soto07}

The choice of thin ($10-100$~nm thick) granular Al films is motivated by their experimental advantages: i) They present a very well defined $T_c$ that minimizes the entanglement between the intrinsic resistivity rounding due to fluctuations and the extrinsic one associated with a possible $T_c$ distribution, mainly near the average $T_c$.\cite{Vidalinh} ii) For temperatures above $\sim3T_c$ the resistivity is almost $T$- and $H$-independent, allowing a reliable determination of the normal-state background. iii) The films properties are determined by the ratio of Al evaporation rate to oxygen pressure and the substrate temperature during evaporation, both of which are easily controlled. As compared to clean systems, their properties are much less dependent on other conditions. Moreover, their superconducting properties present a remarkable correlation with the normal state resistivity.\cite{Levy19} The interest of the $\Delta\sigma$ studies in granular Al films is also enhanced by their implications in other aspects of these superconductors, including the Nernst effect above $T_c$,\cite{Deutscher14} and the understanding of the mechanisms that enhance their $T_c$, a fundamental problem still open.\cite{Pracht16} 

\section{Experimental details}

%
%
\begin{table}[b]
\caption{\label{samples}Growth conditions and superconducting parameters of the studied films, as results from the $\Delta\sigma$ analysis (see the main text for details).}
\begin{ruledtabular}
\begin{tabular}{lccccccc}
 & $d$ & $P_{O_2}$ & $T_c$ & $\rho_n$ & $\xi(0)$ & $\mu_0H_{c2}(0)$ & $\delta$\\
film & (nm) & ($10^{-8}$bar) & (K) & ($\mu\Omega$cm) & (nm) & (T) &  \\
\hline
A & 10 & 1.5 & 2.07 & 163 & 20 & 0.82 & 0.04\\  
B & 10  & 1.0 & 1.90 & 75.2 & 48 & 0.14 & 0.1\\   
C & 100 & 1.5 & 2.07 & 117 & 21 & 0.75 & 0.02\\ 
\end{tabular}
\end{ruledtabular}
\end{table}

\begin{figure}[t]
\begin{center}
\includegraphics[scale=.5]{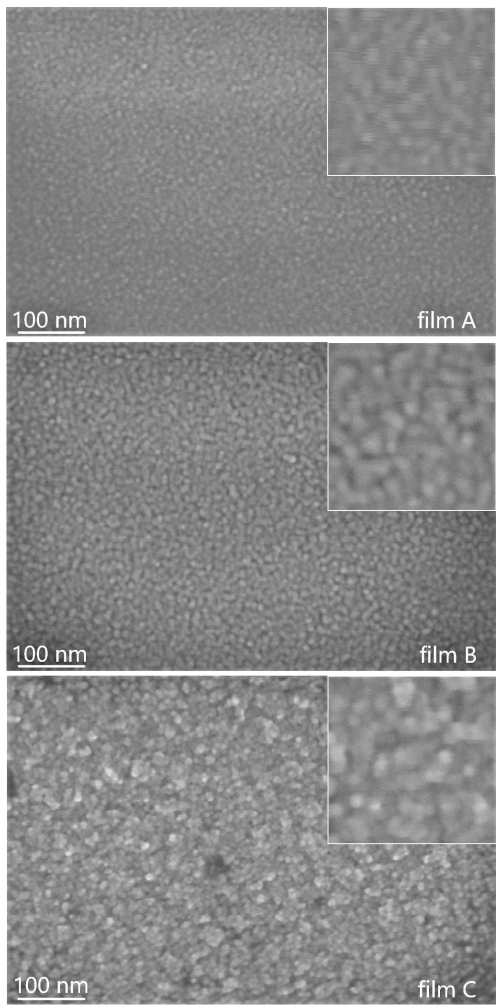}
\caption{Scanning electron micrographs of the samples studied. The detailed views presented in the insets are $100\times100$~nm$^2$ in size.}
\label{sem}
\end{center}
\end{figure}

The films were grown by thermal evaporation of aluminum on $5\times 5$ mm$^2$ thermally oxidized Si substrates at room temperature in a $1-1.5\times10^{-5}$ mbar oxygen pressure. The evaporation rate (0.2 nm/s) was monitored with a quartz balance. The films' microstructure was investigated with a scanning electron microscope (Zeiss FE-SEM Ultra Plus). As it may be seen in Fig.~\ref{sem}, due to the presence of oxygen in the chamber during the evaporation process, they present a granular structure with grain diameters in the 5-10 nm range. In the case of films A and B the nominal thickness is 10 nm and the substrate is essentially covered by a single layer of grains, while in the case of film C (100 nm thickness) they form three-dimensional agglomerates. A summary of the different characteristics of the samples used in this study is presented in Table I.

%
%

The resistivity, $\rho$, was measured with a commercial Physical Property Measurement System (Quantum Design) by using an in-line contacts configuration with an excitation current of 5 $\mu$A, and under perpendicular magnetic fields up to $\sim0.5$~T (large enough to almost quench fluctuation effects above $T_c$). 

\section{Data analysis}

For the sake of clarity this Section focuses on the data for film A. The results for the the other studied films (B and C), that are consistent with the ones for film A, are presented in detail in the Appendix.

\subsection{Analysis of $\Delta\sigma$ for $H=0$ in terms of the conventional AL and MT approaches}

\begin{figure}[t]
\begin{center}
\includegraphics[scale=.6]{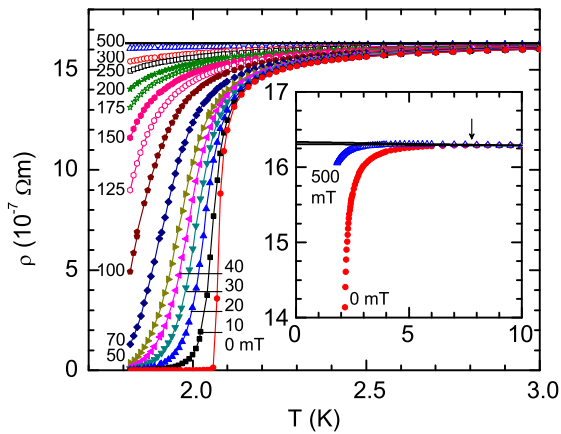}
\caption{(Color online) $T$-dependence of the resistivity of film A around $T_c$ for different perpendicular magnetic fields. Inset: Example for $H=0$ and 500~mT of the procedure used to determine the background contribution by a linear fit above 8~K (indicated by an arrow).}
\label{rho}
\end{center}
\end{figure}

The $\rho(T,H)$ behavior around the superconducting transition is presented in Fig.~\ref{rho}. The $T_c$ value (2.07~K) was estimated as the temperature at which the resistivity falls to 50\% the normal-state background, and the transition width, estimated as $\Delta T=T_c-T(\rho=0)$, is only $\sim 0.02$~K. $T_c$ is two times larger than in bulk Al,\cite{Abeles} an effect early attributed to different finite-size effects,\cite{Parmenter68,Cohen68,Hurault68,Garland68,Dickey68,Pettit76} and that may be strongly affected by the degree of Josephson coupling between neighboring grains.\cite{Mayoh14,Pracht16} The maximum $T_c$ value observed in granular Al films is $\sim3$~K,\cite{Abeles,Pracht16} so the appreciable resistivity rounding extending up to $\sim6$~K (see the inset of Fig.~\ref{rho}) can only be attributed to fluctuations. Moreover, as we will see below, the coherence length amplitudes of the films studied in this work are much larger than the grains average diameter (see Table I). This prevents the appearance of a $T_c$ distribution due to the different grain sizes, and of the associated percolative effects near the average $T_c$. The homogeneous nature of these materials is also supported by the fact that the energy gaps determined from tunneling experiments in similar granular Al films present no detectable broadening in spite of the grains size distribution.\cite{Cohen68}

%
%

\begin{figure}[t]
\begin{center}
\includegraphics[scale=.6]{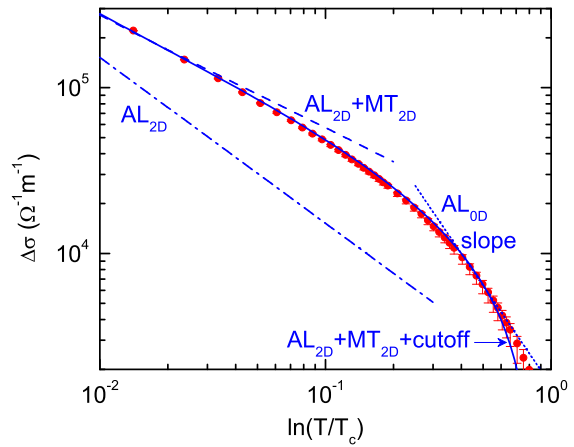}
\caption{(Color online) $\varepsilon$-dependence of $\Delta\sigma$ for $H=0$. The dot-dashed line is the prediction of the 2D-AL approach. The agreement is improved by including the MT contribution (dashed line) but only at low $\varepsilon$. The introduction of a total energy cutoff (solid line) extends the applicability to the high-$\varepsilon$ region. For comparison, the slope $-2$ corresponding to 0D fluctuations is indicated with a dotted line in the high-$\varepsilon$ region. See the main text for details.}
\label{log}
\end{center}
\end{figure}

The paraconductivity was obtained from the data in Fig.~\ref{rho} through $\Delta\sigma(T)=1/\rho(T)-1/\rho_B(T)$, where the almost constant background resistivity $\rho_B(T)$ (solid line) was obtained for each applied field by a linear fit to data above 8~K. The $\Delta\sigma$ dependence on the reduced temperature for $H=0$ is presented in Fig.~\ref{log} in a log-log scale. The dot-dashed line is the AL prediction for two-dimensional (2D) superconductors, given by,\cite{Tinkham, Skocpol75}
\begin{equation}
\Delta\sigma_{AL}^{2D}(\varepsilon)=\frac{e^2}{16\hbar d}\varepsilon^{-1},
\label{eq:2DAL}
\end{equation}
where $e$ is the electron charge, $\hbar$ the reduced Planck constant, and $d=10$~nm the film thickness. As it may be seen, there is a strong discrepancy with the experimental data in both amplitude and critical exponent. However, a good agreement is obtained at low reduced temperatures (dashed line) by also taking into account the anomalous MT contribution for 2D superconductors,\cite{Skocpol75} 
\begin{equation}
\Delta\sigma_{MT}^{2D}(\varepsilon)=\frac{e^2}{8\hbar d}\frac{1}{\varepsilon-\delta}\ln\left(\frac{\varepsilon}{\delta}\right),
\label{MTH0}
\end{equation}
where $\delta$ is the relative shift in the transition temperature due to pair-breaking interactions. The $\delta$ value resulting from the fit for $\varepsilon\alt 0.03$ is $\delta=0.05$, close to the value found in Al films with a similar sheet resistance ($\sim160\;\Omega/\Box$).\cite{Kajimura71,Crow72} The disagreement in the high-$\varepsilon$ region could be interpreted in terms of a 2D-0D dimensional transition associated to the decoupling of the Al grains.\cite{Deutscher77} In fact, in the $\varepsilon$-region bounded by $0.4\stackrel{<}{_\sim}\varepsilon\stackrel{<}{_\sim}0.7$, $\Delta\sigma$ approaches the 0D-AL critical exponent (see Fig.~\ref{log}). However, such a behavior could also be attributed to an overestimation of the statistical weight of high-energy fluctuation modes (with energies of the order of $\hbar^2/2m^*\xi^2_0$, where $m^*$ and $\xi_0$ are the pairs effective mass and size), that may be corrected through the introduction of a cutoff in the energy of the fluctuation modes.\cite{Vidal02} In what follows, this procedure will be applied to the calculus of $\Delta\sigma_{AL}^{2D}$ and $\Delta\sigma_{MT}^{2D}$ to extend its applicability to the high-$\varepsilon$ region. 

\subsection{Generalization of $\Delta\sigma_{AL}$ to high-$\varepsilon$ and -$h$}

A total-energy cutoff was used in Ref.~\onlinecite{Rey13} to extend the Ginzburg-Landau (GL) expression for $\Delta\sigma_{AL}^{3D}$ to the high-$\varepsilon$ and -$h$ regions. The same procedure may be used to derive $\Delta\sigma_{AL}^{2D}$, just by taking into account that for 2D-materials the spectrum of the fluctuations is frustrated along the perpendicular direction (the corresponding component of the fluctuations wavevector, $q_z$ is bounded by $-\pi/d$ and $\pi/d$, and verifies $\xi(0) q_{z}\ll1$). When applied to Eqs.~(B.17) and (B.18) of Ref.~\onlinecite{Rey13}, this leads to\cite{Rey19}
\begin{equation}
\Delta \sigma_{AL}^{2D}(\varepsilon,h,c)=
\frac{e^2   }{32 \hbar d}
\frac{1}{h}
\left[
\psi^{1}\left(\frac{\varepsilon+h}{2h}
\right)-
\psi^{1}\left(\frac{c+h}{2h}
\right)
\right].
\label{ALHc}
\end{equation}
Here $\psi^n$ is nth derivative of the digamma function, and $c$ is the cutoff constant, that corresponds to the $\varepsilon$-value at which $\Delta\sigma$ vanishes. In the zero-field limit ({\it i.e.} for $h \ll \varepsilon,c$), Eq.~(\ref{ALHc}) reduces to
\begin{eqnarray}
\Delta \sigma_{AL}^{2D}(\varepsilon,c)=
\frac{e^2  }{16 \hbar d}\left(\frac{1}{\varepsilon}-\frac{1}{c}\right)\, \, ,
\label{sigma2dener}
\end{eqnarray}
that in absence of cutoff (i.e., $c\to\infty$) leads to Eq.~(1).

%
%

%
%
\begin{figure*}[t]
\begin{center}
\includegraphics[scale=.65]{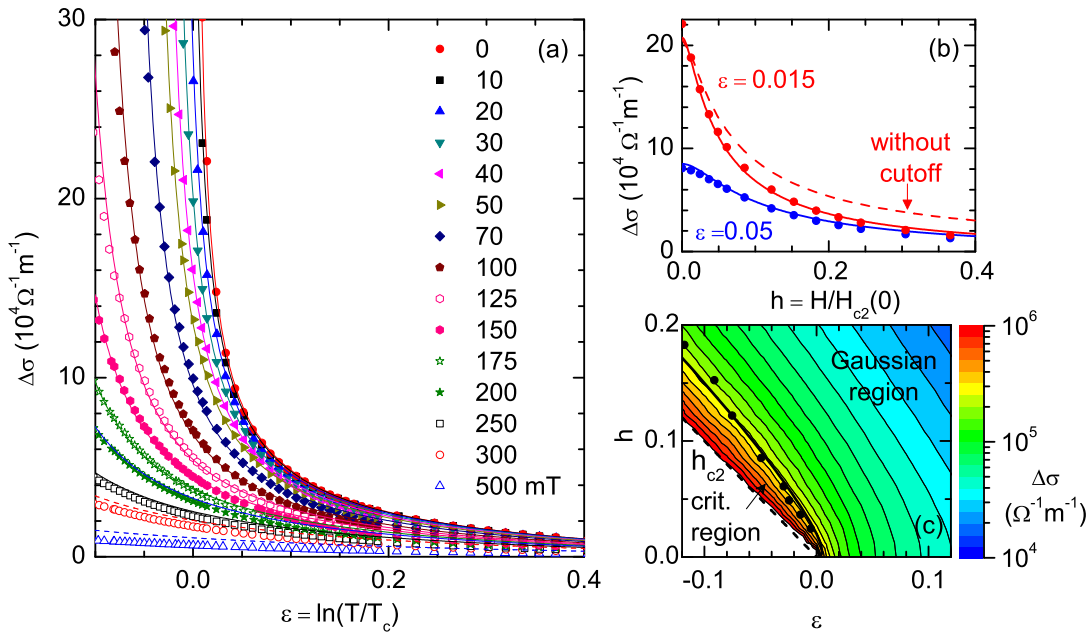}
\caption{(Color online) a) $T$-dependence of $\Delta\sigma$ for film A under different perpendicular magnetic fields. The lines are the best fit of the AL$_{2D}$+MT$_{2D}$ approach under a total-energy cutoff, Eq.~(\ref{ALHc})+Eq.~(\ref{2DMThc}), with $H_{c2}(0)$ as the only free parameter. b) $h$ dependence of $\Delta\sigma$ for selected $\varepsilon$ values. The solid lines are Eq.~(\ref{ALHc})+Eq.~(\ref{2DMThc}) evaluated with the same parameters as in a). If the cutoff is not introduced (dashed line) there is an appreciable disagreement in the high-$h$ region. c) $h-\varepsilon$ phase diagram indicating the $\Delta\sigma$ amplitude (color scale). The data points represent the low-$T$ applicability limit for the theoretical approach used in a). They were obtained as the temperatures at which the theory overestimates by 10\% the experimental data for each $H$. These points agree with the onset of the critical fluctuation region as evaluated from the $H$-dependent Ginzburg criterion (thick solid line, Eq.~(\ref{criterion})). The dashed line is the reduced upper critical field. See the main text for details.}
\label{linH}
\end{center}
\end{figure*}

\subsection{Procedure to introduce a total-energy cutoff in the anomalous MT contribution}

In what concerns the anomalous MT term,\cite{Mori09} the quasiparticles resulting from the decay of a fluctuation of momentum ${\bf q}$ give rise to an extra contribution to the conductivity that should be proportional to the superfluid density, $n_s(q,\varepsilon)\propto[\varepsilon+\xi^2(0){\bf q}^2]^{-1}$, and to a diffusion time $1/Dq^2$ (where $D$ is the diffusion constant) unrelated to the superconducting relaxation time $\tau_s(q,\varepsilon)$.\cite{Patton72} A detailed calculation leads to,\cite{Thompson}
\begin{equation}
\Delta\sigma_{MT}=\frac{\pi e^2}{2\hbar V}\sum_{\bf q}\frac{1}{[{\bf q}^2+\delta/\xi^2(0)][\varepsilon+\xi^2(0){\bf q}^2]}
\label{MT}
\end{equation}
($V$ is the sample volume), that corresponds to Eq.~(21) of Ref.~\onlinecite{Thompson} with the variable change ${\bf q}^2\to{\bf q}^2+\delta/\xi^2(0)$. This transformation accounts for the low momentum cutoff, ${\bf q}^2_{min}=\delta/\xi^2(0)$, to remove the divergence in $\Delta\sigma_{MT}$ occurring in low dimensional systems, allowing to start the sums and integrals in momentum space at ${\bf q}=0$ instead of cutting their low limit. A problem now arises when attempting to apply the total energy cutoff to this equation. While the term associated to $n_s(q,\varepsilon)$ has to be cut off, no restriction has to be imposed to the relaxation time of the quasiparticles. 
To solve these difficulties, let us note that the total-energy cutoff forbids the existence of fluctuations above $\varepsilon = c$, and that $\Delta\sigma_{MT}(\varepsilon,c)$ should vanish at this reduced temperature analogously to the direct contribution. Thus, instead of directly cutting the sum over ${\bf q}$ in Eq.~(\ref{MT}), we have calculated $\Delta\sigma_{MT}(\varepsilon,c)$ by subtracting to $n_s(q,\varepsilon)$ the superfluid density at $\varepsilon =c$,
\begin{equation}
n_s(q,\varepsilon)\longrightarrow n_s(q,\varepsilon,c)=n_s(q,\varepsilon)-n_s(q,c),
\end{equation}
which directly leads to 
\begin{equation}
\Delta\sigma_{MT}(\varepsilon,c)=\Delta\sigma_{MT}(\varepsilon)-\Delta\sigma_{MT}(c)
\label{MTcutoff}
\end{equation}
Following this procedure we reproduce the basic features of the inclusion of a total-energy cutoff (mainly, the vanishing of the fluctuation effects at $\varepsilon=c$), but acting only in the superconducting part of the MT-term. It is worth noting that, when calculating direct contributions to fluctuation effects, this procedure leads to the same results as the direct limitation of the sums over $q$ by the total energy cutoff condition. 

\subsection{Generalization of $\Delta\sigma_{MT}$ to high-$\varepsilon$ and -$h$}

According to Eqs.~(\ref{MTH0}) and (\ref{MTcutoff}), in absence of an external magnetic field the anomalous MT contribution under a total-energy cutoff is given by
\begin{equation}
\Delta\sigma_{MT}^{2D}(\varepsilon,c)=\frac{e^2}{8\hbar d}\left[\frac{\ln(\varepsilon/\delta)}{\varepsilon-\delta}-\frac{\ln(c/\delta)}{c-\delta}\right]
\label{MTH0c}
\end{equation}
As it may be seen in Fig.~\ref{log}, $\Delta\sigma_{AL}^{2D}(\varepsilon,c)+\Delta\sigma_{MT}^{2D}(\varepsilon,c)$ (solid line) now fits the measured $\Delta\sigma$ up to reduced temperatures as high as $\varepsilon\approx0.7$, where the uncertainty associated to the background determination (error bars) is already important. The fitting parameters are $\delta=0.04$ (close to the value obtained by fitting the expression without a cutoff) and $c=0.85$, a value near the one that may be expected for dirty superconductors.\cite{Vidal02}

In what follows we obtain an expression for $\Delta\sigma_{MT}^{2D}$ to analyze the paraconductivity in presence of a perpendicular magnetic field (Fig.~\ref{linH}). Our starting point is Eq.~(\ref{MT}) in the 2D limit, i.e., for $\xi(0)q_z\ll1$. The magnetic field transforms the in-plane spectrum of the fluctuations into the one of a charged particle in a magnetic field.\cite{Landau} Subsequently, one must replace $q_{xy}^2$ by $(n+1/2)4e\mu_0H/\hbar$ (here $\mu_0$ is the vacuum magnetic permeability and $n$ the Landau-level index), and introduce the Landau degeneracy factor $e\mu_0HS/\pi\hbar$, where $S$ is the film surface. An example of the application of this procedure to the calculus of the AL term in the presence of a magnetic field may be seen, for instance, in Ref.~\onlinecite{Rey13}. This leads to
\begin{equation}
\Delta\sigma_{MT}^{2D}=\frac{e^2}{4\hbar d} \sum_{n=0}^{\infty}{\frac{1}{(2n+\delta/2h+1)(\varepsilon+h(2n+1))}}.
\label{eq:2DMThsuma}
\end{equation}
After summing and taking into account Eq.~(\ref{MTcutoff}) to introduce the total-energy cutoff, it results
\begin{eqnarray}
\Delta\sigma_{MT}^{2D}&&(\varepsilon,h,c)=\nonumber\\
&&\frac{e^2}{8\hbar d}\left\{\frac{1}{\varepsilon-\delta}\left[\psi^0\left(\frac{\varepsilon+h}{2h}\right)-\psi^0\left(\frac{\delta+h}{2h}\right)\right]-\right.\nonumber\\
&&\left.\frac{1}{c-\delta}\left[\psi^0\left(\frac{c+h}{2h}\right)-\psi^0\left(\frac{\delta+h}{2h}\right)\right]\right\}.
\label{2DMThc}
\end{eqnarray}
This expression reduces to Eq.~(\ref{MTH0c}) for $h\ll\varepsilon,\delta$, and includes only one additional free parameter, $H_{c2}(0)$, that is implicit in $h$. Note also that for $\varepsilon,\delta,h\ll c$, Eq.~(\ref{2DMThc}) reduces to the $\Delta\sigma_{MT}^{2D}(\varepsilon,h)$ expression without cutoff (see Ref.~\onlinecite{Glatz11}).

\subsection{Analysis of $\Delta\sigma$ for finite $H$ in terms of the generalized AL and MT approaches}

The solid lines in Fig.~\ref{linH}(a) are a fit of Eq.~(\ref{ALHc})+Eq.~(\ref{2DMThc}) by using the above $\delta$ and $c$ values, letting $H_{c2}(0)$ as the \textit{only} free parameter. The data included in the fit correspond to magnetic fields up to 0.2~T and reduced temperatures down to $\varepsilon=0$, although an excellent agreement is observed well beyond those limits (see below). The resulting $\mu_0H_{c2}(0)$ is 0.82~T, which corresponds to a coherence length amplitude of $\xi(0)=[\phi_0/2\pi\mu_0H_{c2}(0)]^{1/2}=20$~nm. This value is larger than the average grains' diameter, indicating that they are strongly coupled. It is also larger than the film thickness ($\sim10$~nm), consistently with the observed 2D behavior. As $\xi(T)$ is always larger than $\xi(0)$, it is expected that no dimensional transition will take place upon increasing the temperature above $T_c$, in particular the 2D-0D transition that would be associated to a decoupling of the grains (although this effect could still be present in films with a smaller $\xi(0)$).

A detail of the field dependence of $\Delta\sigma$ for two selected $\varepsilon$ values is presented in Fig.~\ref{linH}(b). The solid lines are Eq.~(\ref{ALHc})+Eq.~(\ref{2DMThc}) evaluated with the same parameters as in Fig.~\ref{linH}(a). An excellent agreement is observed up to $h\approx0.4$ (i.e., $\mu_0H\approx 0.3$~T). For comparison, the dashed line for $\varepsilon=0.015$ was obtained by using the same approach but without cutoff (i.e., by setting $c\to\infty$). While a good agreement with the data is obtained for $H\to0$ (as expected after the analysis presented in Fig.~\ref{log}), a notable disagreement is obtained in the finite field region. This shows that the introduction of a total-energy cutoff is not only needed to explain the data at high reduced temperatures, but also at high reduced magnetic fields, where the theory also overestimates the contribution of the high-energy modes.\cite{Tinkham,Skocpol75}

The theory agrees with the observed $\Delta\sigma$ down to the data points in the $h-\varepsilon$ phase diagram presented in Fig.~\ref{linH}(c). This limit is consistent with the $H$-dependent \textit{Ginzburg criterion} for the onset of the critical fluctuation region in 2D superconductors (solid line),\cite{crit}
\begin{equation}
T/T_c=1-h-\sqrt{2k_Bh/\xi^2(0)d\Delta c},
\label{criterion}
\end{equation}
as evaluated by using the $\xi(0)$ value obtained above, and a specific heat jump at $T_c$ of $\Delta c=345$~J/m$^3$K, a value in reasonable agreement with the one obtained in similar granular Al films.\cite{DeutscherC}

\section{Conclusions}

The present results show that the seemingly anomalous $\Delta\sigma$ behavior of granular Al thin films at high-$\varepsilon$, attributed in similar films to dimensional transitions, may be explained by introducing a total energy cutoff in the fluctuation spectrum in both the direct AL term and the anomalous MT contribution. Such cutoff is also crucial to describe at a quantitative level the high-$h$ behavior, where the theoretical approaches also overestimate the contribution of the short wavelength fluctuations. These results could thus help to elucidate long standing but still open questions concerning the Al thin films, including the mechanism that enhance their $T_c$. In particular, it has been recently proposed that the Josephson coupling between neighboring grains weakens the quantum confinement in each grain, and should also weaken the associated $T_c$ enhancement. In our films we find $\xi(0)\sim 20-48$~nm, much larger than the grains size ($5-10$~nm), indicating a strong inter-grain coupling. In spite of that, $T_c\sim 1.9-2.1$~K, still as high as two times the bulk value and almost independent on $\xi(0)$.
Our results suggest that the superconductivity in these materials is homogeneous in nature, and that the grains size distribution does not seem to play a relevant role. This questions a purely confinement mechanism for the $T_c$ enhancement, which still remains an open issue. It would be interesting to extend the present study to other superconductors and experimental conditions (e. g., other grain sizes and/or dimensionalities).

\begin{acknowledgments}
This work was supported by FEDER/Ministerio de Ciencia, Innovaci\'on y Universidades – Agencia Estatal de Investigaci\'on (project FIS2016-79109-P), Xunta de Galicia (grant ED431C 2018/11), Xunta de Galicia and FEDER (network ED431D 2017/06 and strategic group ED431E 2018/08), and by the CA16218 Nanocohybri COST Action. Authors would like to thank the use of RIAIDT-USC analytical facilities.
\end{acknowledgments}

\appendix

\section{Experimental results and analysis of the other studied films}

To complement the results on film A described in detail in the main text, here we summarize the experimental results and analysis of films B (grown under a different oxygen pressure), and C (with a larger thickness), that confirm the applicability of our theoretical approach and support the results obtained in film A.

\subsection{Film B}

\begin{figure}[t]
\begin{center}
\includegraphics[scale=.45]{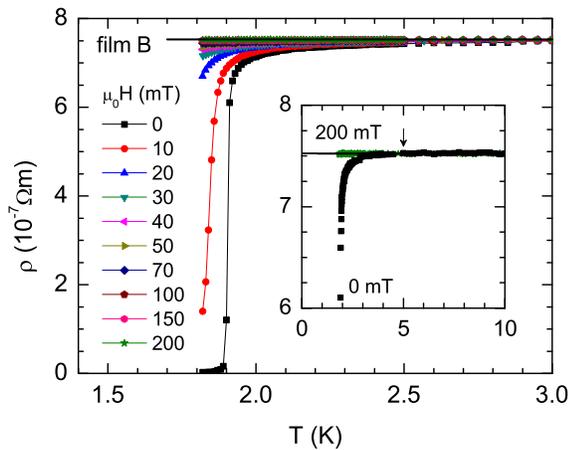}
\caption{(Color online) $T$-dependence of the resistivity of film B around $T_c$ for different $H$. Inset: Example for $H=0$ and 200~mT of the procedure used to determine the background contribution (solid lines) by a linear fit above 5~K.}
\label{rhoB}
\end{center}
\end{figure}

This film was grown under a oxygen pressure 33\% smaller than film A ($10^{-8}$~bar). The temperature dependence of its resistivity around $T_c$ under different perpendicular magnetic fields is presented in Fig.~\ref{rhoB}. The normal state resistivity and the $T_c$ value (1.9~K) are slightly smaller than for film A (see Table~I), probably as a consequence of a better coupling between the Al grains. The background contribution was obtained for each applied field by a linear fit above $T_{onset}=5$~K, the temperature above which fluctuation effects are below the noise level. Examples for $H=0$ and 200~mT are presented as solid lines in the inset of Fig.~\ref{rhoB} (both lines are indistinguishable in this scale).

%
%

\begin{figure}[t]
\begin{center}
\includegraphics[scale=.45]{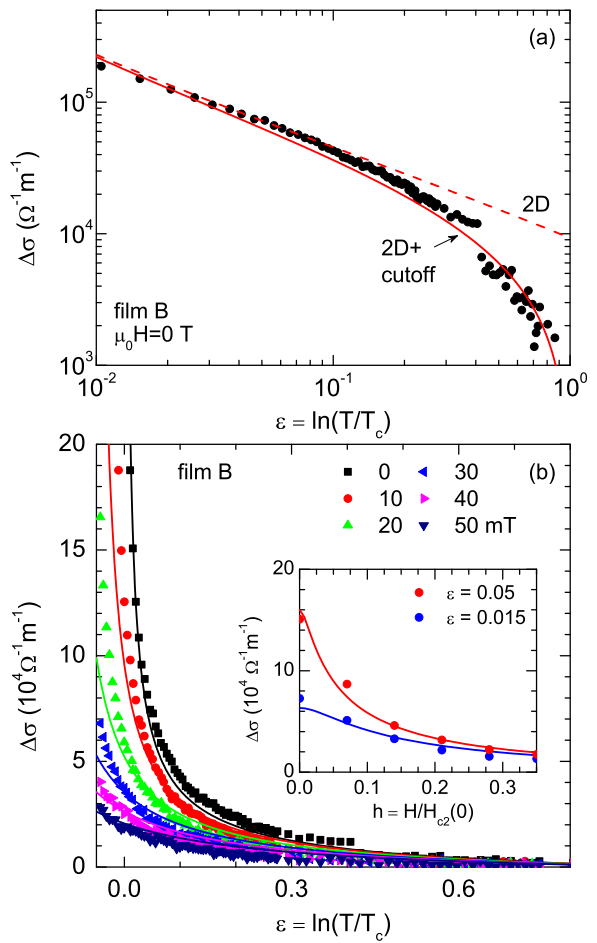}
\caption{(Color online) Comparison of the experimental $\Delta\sigma$ for film B with the 2D AL+MT approach under a total-energy cutoff (solid lines). The dashed line in (a) is the same approach without a cutoff (with $c\to\infty$). The inset in (b) illustrates the $h$-dependence of $\Delta\sigma$ near $T_c$. See the main text for details.}
\label{flucB}
\end{center}
\end{figure}

The paraconductivity for $H=0$ is presented in Fig.~\ref{flucB}(a). The solid line is a fit of $\Delta\sigma_{AL}^{2D}(\varepsilon,c)+\Delta\sigma_{MT}^{2D}(\varepsilon,c)$ [Eq.~(4)+Eq.~(8)], with $c=\ln(T_{onset}/T_c)\approx 1$ and $\delta$ as the only free parameter. The agreement is good in all the accessible $\varepsilon$-region above $\varepsilon\approx 10^{-2}$, and the resulting $\delta$ value ($\sim 0.1$, compiled in Table I), is within the values found in Al films with a similar sheet resistance, see Refs.~\onlinecite{Kajimura71,Crow72}. Just for comparison, the same theoretical approach but without a cutoff (i.e., with $c\to\infty$) is presented as a dashed line. As it may be seen, the agreement in this case extends only up to $\varepsilon\approx 0.1$.

%
%

The $\varepsilon$-dependence of $\Delta\sigma$ under different applied magnetic fields is presented in Fig.~\ref{flucB}(b). The solid lines are the best fit of $\Delta\sigma_{AL}^{2D}(\varepsilon,h,c)+\Delta\sigma_{MT}^{2D}(\varepsilon,h,c)$ [Eq.~(3)+Eq.~(10)], with the above $T_c$, $c$ and $\delta$ values, and with $H_{c2}(0)$ (that is implicit in $h$) as the only free parameter. The agreement is good down to $\varepsilon\approx0$, near the onset of the critical fluctuation region. The resulting $\mu_0H_{c2}(0)$ value (0.14~T) leads to a coherence length amplitude of $\xi(0)=48$~nm, significantly larger than the superconducting grains. This suggests that the grains are strongly coupled, which is consistent with the observed 2D behavior. The $h$ dependence of $\Delta\sigma$ is illustrated in the inset for some $\varepsilon$ values. The lines are the 2D AL+MT approach under a cutoff, evaluated with the same parameters as in the main figure.

\subsection{Film C}

\begin{figure}[t]
\begin{center}
\includegraphics[scale=.45]{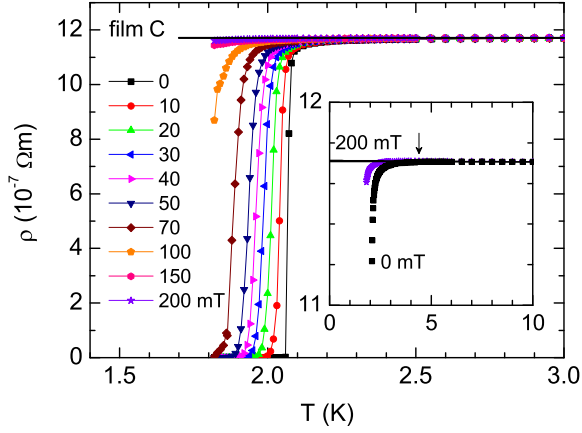}
\caption{(Color online) $T$-dependence of the resistivity of film C around $T_c$ for different $H$. Inset: Example for $H=0$ and 200~mT of the procedure used to determine the background contribution (solid lines) by a linear fit above 4.4~K.}
\label{rhoC}
\end{center}
\end{figure}

This film was grown under the same oxygen pressure ($10^{-8}$~bar) as film A, but is 10 times thicker (100~nm). The temperature dependence of its resistivity around $T_c$ under different perpendicular magnetic fields is presented in Fig.~\ref{rhoC}. As expected, the $T_c$ value (2.07~K) is the same as for film A (see Table I), but the normal state resistivity is slightly smaller, probably as a consequence of its three-dimensional microstructure. The background contribution was obtained for each applied field by a linear fit above $T_{onset}=4.4$~K, the temperature above which fluctuation effects vanish. Examples for $H=0$ and 200~mT are presented as solid lines in the inset of Fig.~\ref{rhoC} (both lines are indistinguishable in this scale).

%
%

%
%
\begin{figure}[t]
\begin{center}
\includegraphics[scale=.45]{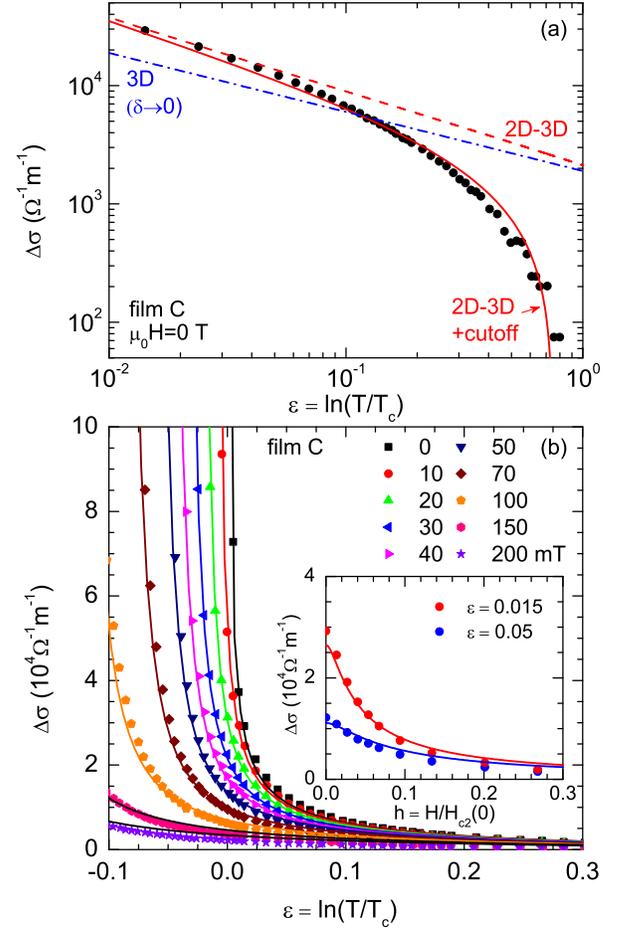}
\caption{(Color online) Comparison of the experimental $\Delta\sigma$ for film C with the 2D-3D AL+MT approach under a total-energy cutoff (solid lines). In (a), the dashed line is the same approach without a cutoff (with $c\to\infty$), that illustrates the disagreement at high-$\varepsilon$, and the dot-dashed line is the best fit of the conventional 3D AL+MT approach. The inset in (b) illustrates the $h$-dependence of $\Delta\sigma$ near $T_c$. See the main text for details.}
\label{flucC}
\end{center}
\end{figure}

In this sample it is possible to appreciate the $T_c$ shift with $H$. This allows to estimate the coherence length from the GL relation 
\begin{equation}
\xi^2(0)=\frac{\phi_0}{2\pi T_c|dH_{c2}/dT|}.
\end{equation}
By using a 10\% (90\%) criterion on the normal-state resistivity to determine $T_c(H)$ it is obtained $\xi(0)=18~(21)$~nm. As this value is $\sim5$ times smaller than the film thickness, it is expected that the fluctuation effects are not 2D in this case. However, the $\varepsilon$-dependence of $\Delta\sigma$ for $H=0$ (Fig.~\ref{flucC}(a)) cannot be explained even at low-$\varepsilon$ by the conventional $AL+MT$ 3D approach (see, e.g., Ref.~\onlinecite{Thompson}) evaluated with $\xi(0)=20$~nm and $\delta=0$ (dot-dashed line). The use of a finite $\delta$ value increases the disagreement. Thus, we have compared the data with a generalization of the AL+MT approach to superconductors with an intermediate 2D-3D dimensionality. This may be easily obtained from the 2D approach presented above by following the procedure described in Ref.~\onlinecite{Thompson}. The result for the AL contribution under a total-energy cutoff is
\begin{eqnarray}
\Delta\sigma_{AL}^{2D-3D}&&(\varepsilon,h,c)=\nonumber\\
&&\frac{e^2}{32\hbar d}\frac{1}{h}\sum_{n=0}^{n_{max}}\left[\psi^1\left(\frac{\varepsilon+h+\xi^2(0)(n\pi/d)^2}{2h}\right)\right.\nonumber\\
&&-\left.\psi^1\left(\frac{c+h+\xi^2(0)(n\pi/d)^2}{2h}\right)\right],
\end{eqnarray}
where $n_{max}=d\sqrt{c-\varepsilon}/\pi\xi(0)$, $d$ is the film thickness, $c$ the cutoff constant, and $\psi^n$ the $n$-th derivative of the digamma function. The anomalous MT contribution in the absence of a cutoff is given by 
\begin{eqnarray}
\Delta\sigma_{MT}^{2D-3D}&&(\varepsilon, h)=\nonumber \\
&&\frac{e^2}{8\hbar d} \frac{1}{\varepsilon-\delta} \sum_{n=0}^{\infty}\left[\psi^0\left(\frac{\varepsilon+h+\xi^2(0)(n\pi/d)^2}{2h}\right)\right.
\nonumber \\
&&-\left.\psi^0\left(\frac{\delta+h+\xi^2(0)(n\pi/d)^2}{2h}\right)\right],
\end{eqnarray}
and the total-energy cutoff may be introduced by using the prescription described in Section III.C,
\begin{equation}
\Delta\sigma_{MT}(\varepsilon,h,c)= \Delta\sigma_{MT}(\varepsilon,h)-\Delta\sigma_{MT}(c,h).
\end{equation}
A detailed calculation of these expressions will be presented elsewhere. It is worth noting that they reduce to the 2D and 3D limits when $d\ll\xi(0)$ and $d\gg\xi(0)$, respectively. 

The solid line in Fig.~\ref{flucC}(a) corresponds to the 2D-3D AL+MT approach under the cutoff, evaluated with $c=\ln(T_{onset}/T_c)\approx 0.75$, $\delta\sim 0.02$ (within the values found in Al films with a similar sheet resistance, see Refs.~\onlinecite{Kajimura71,Crow72}), and $\xi(0)=21$~nm (within the values obtained above from the $T_c(H)$ shift). These values are compiled in Table I. The agreement is now good in all the accessible temperature region above $\varepsilon\approx 10^{-2}$. For comparison, the same theoretical approach but without a cutoff (i.e., with $c\to\infty$) is presented as a dashed line. As expected the agreement in this case extends only up to $\varepsilon\approx 0.1$. As for the other samples, the coherence length is larger than the superconducting grains, which is consistent with a strong intergrain coupling and with the observed non-0D behavior.

The $\varepsilon$-dependence of $\Delta\sigma$ under different applied magnetic fields is presented in Fig.~\ref{flucC}(b). The solid lines are the 2D-3D AL+MT approach with a total-energy cutoff, evaluated with the same parameters as in Fig.~\ref{flucC}(a). As it may be seen the agreement is excellent down to temperatures slightly below $\varepsilon=0$, where critical fluctuations are expected to be appreciable. The $h$-dependence of $\Delta\sigma$ is illustrated in the inset of Fig.~\ref{flucC}(b) for some $\varepsilon$ values. The lines are the 2D-3D AL+MT approach under a cutoff, evaluated with the same parameters as in the main figure.


\begin{thebibliography}{100}
\bibitem{notaMT}The regular MT contribution, as well as the one arising from the density of states (DOS), can be neglected with respect the AL and anomaluos MT terms in conventional $s$-wave superconductors (see Ref.~\onlinecite{Larkinbook}, sec.~7.3).
\end{thebibliography}
\end{document}